# Innovative Application of Artificial Intelligence Technology in Bank Credit Risk Management

Shuochen Bi [1, a], Wenqing Bao [2, b]

[1] D'Amore-McKim School of Business at Northeastern University Boston, Independent Researcher, MA02115, United States

[2] Americold Logistics, LLC, Atlanta, GA, 30319, United States

[a] bi.shu@northeastern.edu; [b] bao.234@osu.edu

## ABSTRACT

With the rapid growth of technology, especially the widespread application of artificial intelligence (AI) technology, the risk management level of commercial banks is constantly reaching new heights. In the current wave of digitalization, AI has become a key driving force for the strategic transformation of financial institutions, especially the banking industry. For commercial banks, the stability and safety of asset quality are crucial, which directly relates to the long-term stable growth of the bank. Among them, credit risk management is particularly core because it involves the flow of a large amount of funds and the accuracy of credit decisions. Therefore, establishing a scientific and effective credit risk decision-making mechanism is of great strategic significance for commercial banks. In this context, the innovative application of AI technology has brought revolutionary changes to bank credit risk management. Through deep learning and big data analysis, AI can accurately evaluate the credit status of borrowers, timely identify potential risks, and provide banks with more accurate and comprehensive credit decision support. At the same time, AI can also achieve real-time monitoring and early warning, helping banks intervene before risks occur and reduce losses.

## KEYWORDS

Artificial intelligence; Bank credit risk management; Deep learning; Big data analysis; Risk intervention

## 1. INTRODUCTION

Since 2015, FinTech has rapidly become a hot topic in the banking industry and even the entire financial sector due to its unique charm and enormous potential [1]. Financial technology is the perfect combination of finance and technology. It uses financial scenarios as carriers and technological innovation as support, bringing unprecedented changes and opportunities to the traditional banking industry [2]. In this wave led by financial technology, the banking industry is undergoing a profound technological revolution, whose impact is far-reaching and wide-ranging, exceeding the expectations of many people [3]. Liquidity, as an important indicator for measuring bank risk, has always been an important component of bank risk management [4]. With the growth of mixed operations in financial institutions, the transmission mechanism of liquidity risk has also undergone profound changes. Recently, the regulatory dynamics of the Federal Deposit Insurance Corporation (FDIC) on community banks have further highlighted the importance of liquidity risk. FDIC has found that community banks in the United States generally face increased liquidity risk. In response to this risk, FDIC requires community banks to improve their emergency financing plans to ensure that they can quickly and effectively obtain the necessary funds when facing potential liquidity pressures, thereby maintaining the stable operation of the bank [5].

However, liquidity risk is only one of the many financial risks faced by commercial banks [6]. Among these risks, credit risk is undoubtedly the most common and important one. Credit risk, in simple terms, refers to the risk that borrowers or debtors are unable to repay debts or fulfill contractual obligations on time due to various reasons, resulting in losses to creditors or banks [7]. This type of risk is common in financial markets and often has significant uncertainty and unpredictability. Taking the subprime crisis in the United States as an example, this global financial crisis is centered around credit risk. In the subprime crisis, a large number of subprime mortgage defaults led to severe asset losses for financial institutions, which in turn triggered global financial turmoil. Major economies such as the United States, the European Union, and Japan have all been affected, and the global financial order has also been significantly negatively affected. More seriously, this crisis also triggered a sovereign debt crisis in Europe, causing the entire European economy to fall into a long-term downturn. The occurrence of the subprime crisis not only reveals the enormous destructive power of credit risk, but also exposes the shortcomings and deficiencies of traditional banking industry in risk management.

In such a severe financial context, commercial banks must further strengthen their risk management capabilities, especially credit risk management. As one of the core businesses of commercial banks, the level of risk control in credit business is directly related to the stable operation and sustainable growth of banks. Therefore, how to effectively identify, evaluate, and control credit risks has become an urgent issue for commercial banks. In this process, the innovative application of AI technology provides new ideas and means for commercial banks. Through big data analysis and machine learning techniques, AI can help banks more accurately evaluate the credit status of borrowers, timely identify potential risks, and provide scientific and effective risk decision-making support for banks.

## 2. THE APPLICATION OF AI IN BANK CREDIT MANAGEMENT

In various fields such as politics, strategy, and society, the uncertainty factors in today's world are more prominent than ever before [8]. This uncertainty not only exacerbates the complexity of the international situation, but also puts higher demands on the credit management of banks [9]. As one of the core businesses of banks, credit management must continuously improve the accuracy of risk warning and decision-making in this context to ensure the stable operation of banks [10]. In this process, the application of AI technology is particularly crucial. Firstly, from a political perspective, the ever-changing international situation has brought unprecedented challenges to bank credit management. Geopolitical conflicts, rising trade protectionism, and fluctuations in international financial markets may all have a significant impact on the credit assets of banks. Therefore, banks need to use AI technology to monitor and analyze the global political and economic situation in real-time, in order to timely identify potential risks and take corresponding risk response measures. At the strategic level, banks need to develop scientific credit policies and management strategies to cope with the constantly changing market environment. AI technology can help banks more accurately evaluate the credit status of borrowers and predict future credit risks through big data analysis and machine learning algorithms.

At the same time, AI technology can also optimize the credit business process of banks, improve business processing efficiency, and reduce operating costs. From a societal perspective, the public's demand for financial security and stability is increasing. As an important component of the financial system, banks must assume the responsibility of maintaining financial stability. The application of AI technology can help banks better identify and manage credit risks, and prevent the spread and amplification of credit risks within the financial system. In addition, AI technology can also improve the intelligence level of banking services, enhance customer experience, and enhance the competitiveness of banks. With the advancement of computer simulation technology, the use of simulation modeling methods for bank risk warning and decision-making has also begun to emerge. Figure 1 shows the process diagram of using AI for complex system modeling. By constructing a risk warning model based on AI technology, banks can monitor and warn credit risks in real time, identify potential risk points in a timely manner, and take corresponding risk response measures. At the same time, AI technology can also provide scientific support for bank credit decisions, helping banks achieve stable growth of credit business under controllable risks.

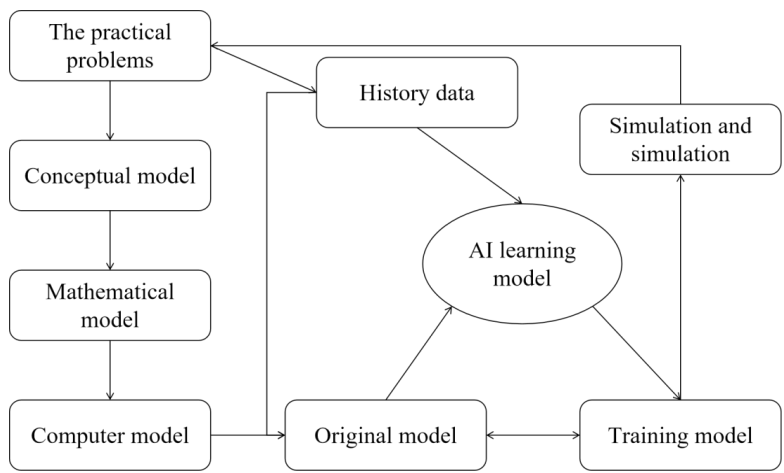

**Figure 1.** Diagram of complex system modeling process using AI

## 3. BANKRUPTCY ANALYSIS OF SILICON VALLEY BANK

Silicon Valley Bank adopted an active deposit absorption and asset allocation strategy during the period of loose liquidity. However, this strategy exposed serious risks when the Federal Reserve's monetary policy shifted towards interest rate hikes, especially in the context of "interest rate inversion". The bank, while absorbing a large amount of low interest corporate deposits, has also allocated long-term limited bond assets, resulting in a prominent mismatch of asset liability terms. When the market environment changes, the cost of debt rises, and the prices of held bond assets plummet, the liquidity of banks quickly depletes and eventually falls into a predicament. Table 1 shows the loan structure of Silicon Valley Bank in 2022, sourced from the annual report of Silicon Valley Bank in 2022.

**Table 1.** Loan Structure of Silicon Valley Banks in 2022

| Loan type | Balance (million US dollars) |
|---|---|
| Global industrial investment fund | 41269 |
| Investor linked loans | 6713 |
| Cash flow linked loans | 1966 |
| Innovative industrial and commercial loans | 8609 |
| Private bank loans | 10477 |
| Commercial real estate loans | 2583 |
| Winery loan | 1158 |
| Other industrial and commercial loans | 1019 |
| other | 433 |
| Salary guarantee loan | 23 |
| LOAN | 74250 |
| Impairment of Assets | -636 |
| Net loans | 73614 |

From the loan structure of Silicon Valley Bank, its asset structure has undergone significant changes in recent years. The proportion of loans has been decreasing year by year, while the proportion of bonds has been increasing year by year. This change in asset structure makes banks more dependent on the performance of the bond market, which is highly volatile. Once the market environment changes, the value of bank assets may be severely affected. In addition, Silicon Valley Bank overly relies on low interest deposits in the PE/VC market for deposit absorption. This type of deposit has high liquidity, and once market liquidity tightens, these deposits may quickly drain, bringing liquidity risk to banks. At the same time, as the Federal Reserve raises interest rates, depositors' demands for interest income are also increasing, which has led to an increase in the cost of debt for Silicon Valley banks and further intensified their operational pressure. Against the backdrop of rapid interest rate hikes, bond prices have plummeted, and Silicon Valley Bank's bond assets are facing significant potential losses. According to statistics, Silicon Valley Bank had already gone bankrupt technically by the end of September 2022. This incident not only brought huge losses to Silicon Valley Bank itself, but also had a profound impact on the entire financial market. The strategy adopted by Silicon Valley Bank during the period of loose liquidity exposed serious risks during the interest rate hike cycle. The mismatch of its asset liability period, unreasonable asset structure, and excessive reliance on low interest deposits in the PE/VC market have all led to its ultimate predicament. This incident also reminds other banks to fully consider changes in the market environment during their operations, allocate assets and liabilities reasonably, and reduce risks.

## 4. COUNTERMEASURES FOR BANK CREDIT RISK MANAGEMENT

Banks must always adhere to the three basic principles of safety, liquidity, and profitability in their business management, which are fundamental to the stable operation of banks. Especially in the context of rapid technological advancements, how banks can utilize new technologies, especially AI, to strengthen credit risk management has become an important issue in front of us. With the help of AI technology, banks can analyze multi-dimensional information such as customer credit records, financial status, and business behavior more deeply, thereby more accurately evaluating customer credit status. Meanwhile, the real-time monitoring function of AI can timely detect the trend of credit risk changes and provide timely risk warnings for banks. In this way, banks can take effective measures to reduce the likelihood of credit losses before risks occur. With the help of AI, banks can more accurately predict and plan the flow of credit funds. Through big data analysis, banks can understand the funding needs and repayment ability of customers, and thus arrange the allocation and recovery of credit funds in a reasonable manner.

In addition, AI can also help banks optimize credit processes, improve business processing efficiency, and ensure that credit funds can be timely and accurately received. Through the application of AI, banks can more scientifically formulate credit policies, optimize credit structure, and improve the quality of credit assets. At the same time, AI can also help banks reduce operating costs, improve operational efficiency, and thereby enhance their profitability. However, in the process of using AI for credit risk management, we must also attach great importance to the protection of customer privacy information and rights. When collecting and using customer information, the principle of minimum necessity must be followed, and excessive collection or abuse of customer information is not allowed. Meanwhile, banks should also establish sound internal control mechanisms to prevent internal personnel from leaking or abusing customer information. In addition, in order to ensure the secure application of AI technology, banks also need to strengthen the security management of technology applications. Through these measures, banks can effectively prevent the risks that may arise from the application of new technologies.

# 5. THE CHALLENGES AND PROSPECTS OF AI TECHNOLOGY IN BANK CREDIT RISK MANAGEMENT

The application of AI technology in bank credit risk management undoubtedly brings many innovative opportunities, but it also comes with a series of challenges. With the digital transformation of banking business, a large amount of sensitive data is generated, stored, and processed. AI technology relies on this data to train models and make predictions, but it also increases the risk of data leakage and misuse. On the one hand, banks need to ensure the security of data, prevent hacker attacks and internal leaks; On the other hand, with the increasing global concern for data privacy, banks also need to comply with increasingly strict data protection regulations.

Compared with traditional risk assessment methods, AI models based on machine learning are often more complex and difficult to explain. This "black box" characteristic may make the results of the model difficult to trust and understand, especially when it comes to major financial decisions. Regulatory agencies and clients may require banks to provide detailed information about the decision-making process to ensure fairness and transparency.

With the widespread application of AI technology in the financial field, ethical and regulatory issues related to it are becoming increasingly prominent. For example, if an AI based credit decision-making model has bias or discrimination, it may lead to unfair loan conditions or denial of service. In addition, regulatory authorities are continuously improving the regulatory framework and standards for AI technology in the financial field. Banks need to closely monitor regulatory developments while complying with existing regulations.

**Table 2.** Challenges and prospects of AI application

| Challenge/Outlook | Describe | Affect |
| --- | --- | --- |
| Data security and privacy protection | Security and privacy protection challenges faced by banks when dealing with sensitive data. | It may lead to data leakage, lack of trust and legal risks. |
| The interpretability and transparency of models | The model based on machine learning is complex and difficult to explain, which affects the credibility and transparency of decision-making. | It may lead to regulatory compliance problems, lack of customer trust and misjudgment risks. |
| Technological updates and iteration speed | It is challenging for banks to keep up with the rapid development of financial technology. | It may lead to backward technology, declining competitiveness and business interruption. |
| Ethical and regulatory issues | Ethical and regulatory concerns caused by the application of AI technology in the financial field. | It may lead to unfair decision-making, regulatory punishment and reputation risk. |
| Intelligent risk management | Using AI technology to improve the intelligent level of risk management. | The efficiency, accuracy and response speed of decision-making can be improved. |
| Comprehensive risk management | Combine technologies such as big data and cloud computing to achieve more comprehensive risk management. | It can identify, evaluate and manage all kinds of risks more deeply. |
| Personalized risk management | Develop accurate and personalized risk management solutions for customers. | Can improve customer satisfaction, loyalty and market competitiveness. |

Despite facing many challenges, the application prospects of AI technology in bank credit risk management are still broad. Table 2 summarizes the challenges and prospects of AI applications. With the continuous advancement of technology and the reduction of costs, more and more banks will be able to adopt advanced AI technology to improve the intelligence level of risk management. AI technology will be combined with other financial technology tools such as big data and cloud computing to achieve more comprehensive and in-depth risk management. By comprehensively analyzing data from multiple sources and types, potential risks and opportunities can be more accurately identified. AI technology will also promote the personalized development of bank risk management. Through in-depth analysis and understanding of customers, banks can develop more personalized risk management plans for each customer.

## 6. CONCLUSIONS

The continuous iterative growth of AI technology is profoundly changing the operational mode and risk management paradigm of the banking industry, promoting the level of decision-making management to new heights. For banks, decision-making is not only related to the efficiency and effectiveness of daily operations, but also the cornerstone of ensuring fund security, maintaining customer trust, and achieving stable growth. In the field of credit risk management, the application of AI technology makes the decision-making process more scientific and accurate. Through deep mining and analysis of massive data, AI can help banks gain a more comprehensive understanding of their customers' credit status and repayment ability, thereby formulating more reasonable credit policies. At the same time, real-time dynamic monitoring and early warning mechanisms enable banks to quickly respond to risk events and reduce losses. With the continuous growth of AI technology, its application in decision management will also be more extensive and in-depth. In the future, banks will be able to utilize higher-level algorithms and models to achieve more intelligent decision-making and further enhance the efficiency of risk management.